\documentclass[runningheads]{llncs}
\usepackage{graphicx}
\usepackage{amsfonts}
\usepackage{color,soul}
\usepackage[ruled]{algorithm2e}
\usepackage{algorithmic}
\usepackage{multirow}
\usepackage{tabularx}
\usepackage{tikz}
\usetikzlibrary{matrix,shapes,arrows,positioning,chains,calc}
\usepackage{adjustbox}
\usepackage{mathtools}
\usepackage{comment}
\usepackage{numprint}
\PassOptionsToPackage{obeyspaces,hyphens}{url}
\usepackage[hyperindex]{hyperref}
\begin{document}
 
\title{QSOR: Quantum-Safe Onion Routing}


	\author{Zsolt Tujner\inst{1,2} \and
Thomas Rooijakkers\inst{1} \and
Maran van Heesch\inst{1} \and
Melek \"{O}nen\inst{2}}
\authorrunning{Z.\ Tujner, T.A.\ Rooijakkers, M.\ van Heesch, M.\ \"{O}nen}
%
\institute{TNO, the Hague, the Netherlands\\
\email{firstname.lastname@tno.nl}\\
\url{https://www.tno.nl/} \and
EURECOM, Sophia-Antipolis, France\\
\email{melek.onen@eurecom.fr}\\
\url{https://www.eurecom.fr/} 
}




\newcolumntype{Y}{>{\centering\arraybackslash}X}

\maketitle
\begin{abstract}

{In this work, we propose a study on the use of post-quantum cryptographic primitives for the Tor network in order to make it safe in a quantum world. With this aim, the underlying keying material has first been analysed. We observe that breaking the security of the algorithms/protocols that use long- and medium-term  keys (usually RSA keys) have the highest impact in security. Therefore, we investigate the cost of quantum-safe variants. These include key generation, key encapsulation and decapsulation. Six different post-quantum cryptographic algorithms that ensure level 1 NIST security are evaluated. We further target the Tor circuit creation operation and evaluate the overhead of the post-quantum variant. This comparative study is performed through a reference implementation based on SweetOnions that simulates Tor with slight simplifications. We show that a quantum-safe Tor circuit creation is possible and suggest two versions - one that can be used in a purely quantum-safe setting, and one that can be used in a hybrid setting.
}
\keywords{Tor, anonymous routing, post-quantum cryptography}
\end{abstract}

\section{Introduction}
\label{introduction}
Nowadays, information available online is expanding in an unforeseen way, a vast amount of data is uploaded and shared through social media, IoT, etc.\ ~\cite{datacreate}. However, this data also attracts unwanted attention and might paint a bad image of some stakeholders. Consider the case of Edward Snowden who put the National Security Agency (NSA) in the spotlight by shedding light on how the American population was wiretapped~\cite{snowden}. When blowing the whistle on such a large scale one would aim to remain anonymous, as this act can negatively affect the career and freedom of the individual. In oppressive regimes, where the freedom of speech is abused, this is even more serious, as any type of negative speech, whistle-blowing or expressing freedom of information may be recognized as an act of treason resulting in severe punishments.

The Onion Router (Tor~\cite{torpaper}) aims to obfuscate the anonymity of its users when accessing or communicating over the Internet. In principle, when using Tor, the messages or website connection requests are sent through a network of relays and after multiple 'hops' reach their destination. So, if Alice wants to send Bob a message, but does not want an eavesdropper to know that she initiated the contact, Alice can use Tor. The cryptographic schemes used today and in Tor are based on hard mathematical assumptions e.g., Discrete Logarithm Problem and integer factorization \cite{KatzLindell}. These schemes are assumed to be secure against classical adversaries, as solving them with the currently known algorithms cost exponential time. However, with a quantum computer solving these problems become feasible.

The transition from current cryptography to post-quantum cryptography needs to be started as soon as possible as quantum computers pose a threat to current and in particular public-key cryptographic algowithms. It is expected that this will have significant effects on IT infrastructure. This is due to the heavier operations quantum-safe cryptography requires for setting up connections~\cite{hybrid}. Furthermore, network load is also expected to increase as message sizes are bound to get larger due to the increased encryption sizes. Tor is a volunteer run network all across the globe and both the people running the nodes and the users connecting are going to experience drawbacks. 

\paragraph{Contributions}
\label{researchQs}
In this work, we investigate the main challenges to build and maintain a quantum-safe Tor network. We first examine the different keying material used in Tor and identify the impact of the compromise of each of them. We observe that the migration towards quantum-safe Tor should start with the update of cryptographic algorithms that involve long-term and medium-term keys such as the identity key. Such a migration naturally results in additional cost in terms of CPU and bandwidth.

In order to evaluate the actual overhead resulting from the shift to post-quantum cryptographic algorithms, we have conducted an experimental study while considering six different post-quantum public-key encryption algorithms that are part of NIST's round 2 submissions\footnote{\url{https://csrc.nist.gov/Projects/post-quantum-cryptography/round-2-submissions}}. The implementation of these algorithms is provided by the Open Quantum Safe library~\cite{oqs}. We particularly analyze the cost of key generation, key encapsulation and key decapsulation, and compare them with the currently used algorithms deriving from the \texttt{RSA} cryptosystem or elliptic curve cryptography. We observe that each implementation comes with different advantages and limitations, and that consequently there is no ideal solution that offers optimal CPU and bandwidth overhead.

We further focus on a Tor network operation, namely circuit building, and compare the cost of the post-quantum variant operation with the original one. This comparative study is performed through a reference implementation based on SweetOnions\footnote{\url{https://github.com/LeonHeTheFirst/SweetOnions}, accessed on 28/11/2019, 11:21am} that simulates Tor with slight modifications. Additionally, similar to~\cite{hybrid,hybridor}, a hybrid implementation combining the use of post-quantum cryptography with classical schemes is also proposed. Such an implementation protects against potential security flaws of quantum-safe schemes due to their recent publications. We show that while the increase in CPU time is acceptable and similar among different implementations, the bandwidth overhead remains significant and the optimal performance is achieved when Sike~\cite{sike} is used.

\section{Background}
\subsection{The Onion Router (Tor)}

\begin{figure}[h!]
  \centering
  \includegraphics[width=\linewidth]{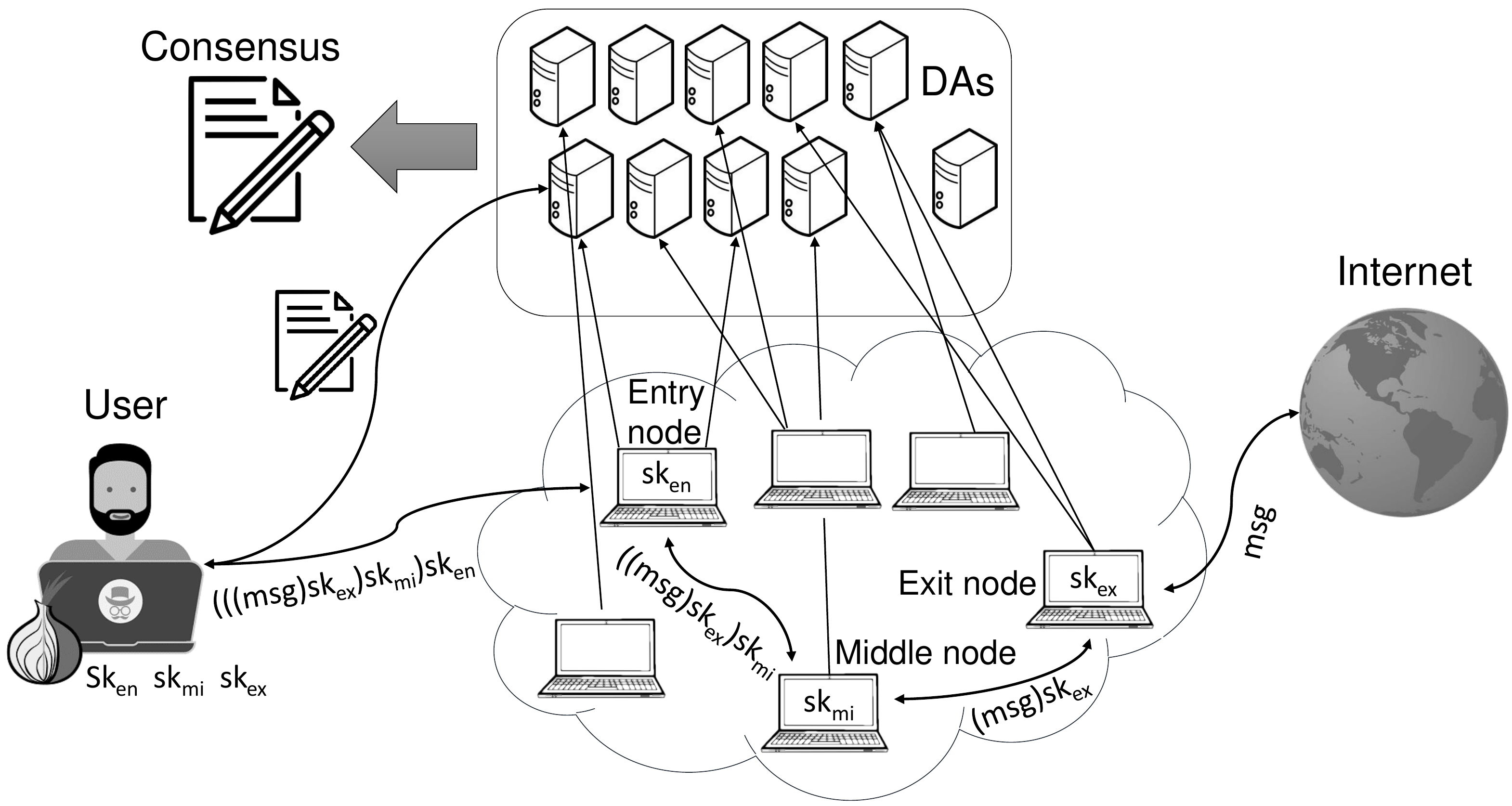}
  \caption{An overview of Tor containing nine directory authorities (DAs), a bridge authority, the consensus document, Tor nodes, the symmetric keys (sk), and the message (msg).}
  \label{fig:torcircuitfig}
\end{figure}

The onion routing network, Tor~\cite{torpaper}, is one of the most popular tools to achieve anonymity for web browsing. When a Tor user accesses a website on the Internet, the traffic encrypted with multiple encryption layers is routed across multiple relays. The use of multiple nodes enroute to the destination helps obfuscate the connection of users and hence achieve anonymity: Each node in the path towards the destination (named a circuit), only has information about the previous node and the next node in the path. Messages are encrypted by the source in a layered fashion whereby each encryption layer is removed by one relay node. Nowadays, Tor counts around \numprint{6000} nodes\footnote{\url{https://metrics.torproject.org/networksize.html}, accessed on 28/11/2019, 4:33pm} and the default number of relay nodes to set up a circuit is three (entry node, middle node, exit node)\footnote{\url{https://trac.torproject.org/projects/tor/wiki/TorRelayGuide}, accessed on 28/11/2019, 4:33pm}. Each node has to communicate information called descriptors towards Directory Authorities, who maintain a state of the network. The Directory Authorities vote on the status of the network to obtain a consensus document. The user connects to one of the Directory Authorities, fetches the consensus document and the Tor software will select a path from the available nodes. The overall Tor framework is illustrated in Figure~\ref{fig:torcircuitfig}.
The security and privacy features that Tor ensures rely on the use of essential cryptographic primitives such as encryption, digital signatures and key exchange. Consequently, each Tor node receives and maintains multiple cryptographic keys for different purposes. Table~\ref{tab:keyusetable} provides an overview of the asymmetric keys used in Tor with their lifetime and functionalities. Long-term keys are used at least for one year, medium-term keys are used for three to twelve months, and short-term keys have a lifespan of minutes to a maximum of one day. 
\begin{table}
    \begin{adjustbox}{max width=\linewidth, center}
    \begin{tabular}{|l|l|l|p{0.5\linewidth}|}
    \hline
        \textbf{Type} & \textbf{Key lifetime} & \textbf{Key name} & \textbf{Function} \\
         \hline
         \multirow{3}{*}{\texttt{RSA}} & long-term & identity key & Establish relay identity, sign documents and certificates. Since the introduction of \texttt{Ed25519}, \texttt{RSA} is only used to establish relay identity.\\
         \cline{2-4}
         & medium-term & onion key & Decrypt cells at circuit creation. Used in \texttt{ntor} and TAP for handshakes. \\
         \cline{2-4}
         & short-term & connection key & Establish TLS channels between nodes.\\
         \hline
         \texttt{Curve25519} & medium-term & handshake key & Handle handshakes in the \texttt{ntor} protocol.\\
         \hline
         \multirow{3}{*}{\texttt{Ed25519}} & long-term & master identity key & Sign medium-term \texttt{Ed25519} key. This key never changes.\\
         \cline{2-4}
         & medium-term & signing key & Replaces \texttt{RSA} identity key to sign documents and certificates. \\
         \cline{2-4}
         & short-term & link authentication key & Authenticate handshakes after a Tor circuit negotiation.\\
         \hline
    \end{tabular}
    \end{adjustbox}
    \caption{Function of \texttt{RSA}, \texttt{Curve25519} and \texttt{Ed25519} keys in Tor.}
    \label{tab:keyusetable}
\end{table} 

\subsection{Post-quantum cryptography}
\label{postquantumcrypto}
The security of the current asymmetric encryption and digital signature standards mostly depend on the hardness of integer factorization (\texttt{RSA}) or discrete logarithm (Diffie-Hellman, Elliptic Curve Discrete Logarithm) \cite{KatzLindell}. As described in~\cite{shorpaper}, such cryptographic schemes can be easily broken in polynomial time when using quantum computers. Hence, researchers are actively developing post-quantum cryptographic solutions to resist quantum attacks \cite{Bernstein2009}.  

In 2016, the National Institute of Standards and Technology (NIST) opened a call for proposals on the topic of quantum-safe cryptographic solutions for new quantum-safe standards~\cite{niststandard}. The first round contained 69 submissions. On January 30, 2019 the candidates for the second round were announced, consisting of 17 asymmetric key encryption and key-establishment algorithms and 9 digital signature algorithms.

The transition to quantum-safe cryptographic schemes is expected to be a lengthy process. The adoption of quantum-safe schemes results in a significant increase in bandwidth and computational cost. Developers adopt the hybrid approach whereby currently used standardized cryptographic schemes are combined with quantum-safe schemes. 

\subsection{Related work}
\label{RelWork}

At the time of writing, there are two papers that consider a quantum-safe Tor network, namely~\cite{hybridor} and~\cite{hybrid}. Both solutions mainly focus on the problem of key exchange. In~\cite{hybridor}, the proposed solution named \texttt{HybridOR} is a customized key exchange protocol. The solution is reported to be computationally more efficient compared to currently used the \texttt{ntor} protocol. \texttt{HybridOR} is assumed to be secure under the ring-Learning With Error (r-LWE) assumption. In~\cite{hybrid}, the focus is also on securely establishing the short-term keys in a quantum-safe fashion. The currently used \texttt{ntor} protocol is modified and is called \texttt{Hybrid}. \texttt{Hybrid} uses a combination of long-term keys generated by Diffie-Hellman key exchange, and short-term keys generated by a quantum-safe scheme \texttt{NTRUEncrypt}. We observe that existing solutions focus on the problem of key exchange only. Furthermore, their performance study only focuses on the use of one particular quantum-safe cryptographic scheme. For example, \texttt{Hybrid} is evaluated using the \texttt{NTRUEncrypt} algorithm only. Therefore there seems to be a lack of comparative study among different quantum-safe cryptographic primitives. 

\section{Migration towards quantum-safe Tor}
\label{PQTor}
In a quantum world, the users of Tor need to have the same security and anonymity guarantees as they currently have in a classical setting. A quantum-safe Tor network should provide users security and anonymity against quantum adversaries whilst preserving the security and anonymity claims against classical adversaries. Current attack scenarios on Tor do not target the cryptography used in Tor, but aim to exploit other potential weaknesses. However, powerful enough quantum computers will pose a new threat to Tor as cryptography becomes vulnerable for abuse by quantum adversaries.

\subsection{Challenges}
 With quantum computing emerging, the cryptography of Tor needs to be changed, as the quantum vulnerability of asymmetric key cryptography will open a new attack surface for quantum adversaries. Introducing post-quantum cryptography to Tor must be done in order to keep cryptographic vulnerabilities off the list of attack surfaces. It is pivotal to introduce quantum-safe cryptography to the keys of the nodes. Table~\ref{tab:attackercapabilitiesasymkeys} explains the attacker capabilities in case the \texttt{RSA}, \texttt{Curve25519} or \texttt{Ed25519} schemes are broken, thereby compromising the keys of the nodes.
 \begin{table}[!htbp]
    \begin{adjustbox}{max width=\textwidth, center}
    \begin{tabular}{|l|p{0.62\textwidth}|}
    \hline
         \textbf{Key} & \textbf{Attacker capability} \\
         \hline
          Identity key (long-term) & Impersonate a node, send spoofed descriptors that are signed by the compromised identity key. \\
         \hline
          Onion key (medium-term) & Read content of Tor cells until the next key rotation. \\
         \hline
          Connection key (short-term) & See encrypted traffic between nodes. \\
         \hline
          Handshake key (medium-term) & Read content of Tor cells when a circuit is created. \\
         \hline
         Master identity key (long-term) & Create a new signing key.\\ \hline
				 Signing key (medium-term) & Sign modified documents and publish them to the directory servers. \\
        \hline
          Link authentication key (short-term) & Authenticate connections that should be not allowed. \\
         \hline
    \end{tabular}
    \end{adjustbox}
    \caption{Attacker capabilities with compromised asymmetric schemes.}
    
    \label{tab:attackercapabilitiesasymkeys}
\end{table}
\subsubsection{Current attack scenarios}
Current attacks on Tor do not target the cryptography, but rather focus on vulnerabilities in Tor related software, hidden services, bridge node discovery, disabling the network, and on generic attacks like timing. A common technique of adversaries is to introduce new nodes to the Tor network, but this is a lengthy process due to the policy of the network. New nodes are even more closely monitored than nodes already in the network for malicious patterns and if such is recognized, they are excluded from the network.

\subsubsection{Cryptographic attacks}
In this section, we consider, attack scenarios on the keys of the nodes that a quantum adversary possesses.

There are four types of keys in Tor (See Table \ref{tab:keyusetable}) and all of these can be compromised by an attacker:
\begin{itemize}
    \item \emph{short-term key}. Compromising a short-term key at an entry node would make an adversary capable to follow the entire circuit from sender to recipient. This would lead to deanonymization of the user. After ending a TLS connection, the keys are renewed. Therefore, an attack on a short-term key is performed during the lifetime of its TLS connection.

    \item \emph{medium-term key}. In case an adversary compromises the short-term key and the medium-term key of a node, the attacker can impersonate this node. Since a node can decrypt one layer of symmetric encryption when the messages are passed through it, the previous and next `hop' in the circuit are learned by the adversary. The attack has to be performed before the rotation of the medium-term and short-term keys.

    \item \emph{long-term key}. The long-term key may also be compromised by an adversary. This would enable the adversary to impersonate the node and send forged descriptors to the directory nodes. Furthermore, it allows the adversary to gain indefinite, full access to the node. Moreover, the adversary sees previous and consecutive `hops' in the circuit with the encrypted cells.

    \item \emph{symmetric key}. Symmetric keys are used to encrypt the data sent between nodes. In the current implementation of Tor, \texttt{AES} 128-bit is used for the symmetric encryption~\cite{torspec}. The key size of this scheme should be doubled in order to remain safe against quantum adversaries. The \texttt{AES} 256-bit scheme is claimed to achieve 128-bit security against quantum adversaries \cite{Bernstein2009}.
\end{itemize}

Compromising the symmetric keys enables an adversary to decrypt layers of encryption and learn the destination of the message; anonymity is at risk. In case the attacker learns nothing but the symmetric keys, the encrypted message must be intercepted before entering the network as the TLS connection will add an extra layer of security. If an adversary does not learn all the symmetric keys, but only a subset, then it cannot fully decrypt the message and thus, the circuit is not fully known, so source and destination remain anonymous.

Current attacks on Tor are carried out with colluding adversaries. If adversaries control the entry and exit nodes in the network, they can share information with each other and as a result deanonymize communicating parties. Colluding adversaries at the entry and at the exit node who have the medium-term keys will both know the middle relay in a circuit. Sharing this knowledge enables them to attempt to deanonymize users, as the users using the common middle node have the greatest probability to be communicating with each other.

\subsection{Migration strategy for quantum-safe Tor}\label{sec:MigStrat}
Considering Table~\ref{tab:attackercapabilitiesasymkeys} and the lifetime of the asymmetric keys, it is most urgent to update the long-term keys to a quantum-safe alternative. Long-term keys remain unchanged for a long time-period. Hence, an adversary has more time to compromise long-term keys. The effects of compromising long-term keys are also greater, as an adversary can thereby impersonate a node. The second most urgent, is updating the medium-term keys based on the available time period for compromising is smaller. Finally, the short-term keys must be considered, even though the attacker has limited time to compromise these keys due to the security restrictions of Tor. Furthermore, short-term keys are used with TLS, and there are works on making TLS quantum-safe~\cite{postquantumtls}. We do stress that it is crucial to update every asymmetric scheme to a quantum-safe alternative in order to enforce the security and anonymity claims of Tor.

Lastly, we note that the symmetric keys must be updated to \texttt{AES} 256 bits to prevent `store now, decrypt later'-attacks and ensure that users of Tor maintain life-long anonymity.

\section{Impact of post-quantum cryptography on Tor}
In this section, we investigate the impact that post-quantum cryptography might have on the Tor network, when following the suggested migration strategy in Section \ref{sec:MigStrat}. The impact of migrating all asymmetric cryptography to a quantum-safe alternative has an impact on the performance (both computational and network) and reliability of Tor. We focus, in particular, on the key exchanges as updating these has the greatest effect on the overall performance and reliability of Tor.

\subsection{Benchmarking post-quantum cryptography}
\label{sec:benchmark}
We benchmark the post-quantum cryptography schemes that have been implemented in the Open Quantum Safe library~\cite{oqs}.

\subsubsection{Open Quantum Safe library}The Open Quantum Safe library contains multiple implementations of post-quantum secure key encapsulation and signature schemes.
The schemes reach NIST security levels 1 to 5, however we only tested the schemes that achieve level 1 NIST security. The tested schemes are listed in Table~\ref{tab:pqkeys}.

\subsubsection{System setup}
\label{systemsetup}
For the experiment, local and virtual environments are both used. The technical specification of the notebook used for the local experiments is Dell Latitude E7240, with Intel Core i5-4310U CPU @ 2.00 - 2.60GHz processor, 8 GB RAM, Samsung SSD SM841N mSATA 128 GB for storage and Windows 10 Enterprise 64-bit operating system. Furthermore, an Ubuntu 18.04 LTS subsystem was installed. In order to emulate the Tor network, 6 virtual machines were used with Intel Core Processor (Broadwell) @ 2.4 GHz processors, 60 GB storage, a virtual network adapter, and Linux version 4.15.0 operating system.


\subsubsection{Methods and results}
\begin{figure}[h!]
    \includegraphics[width=\linewidth]{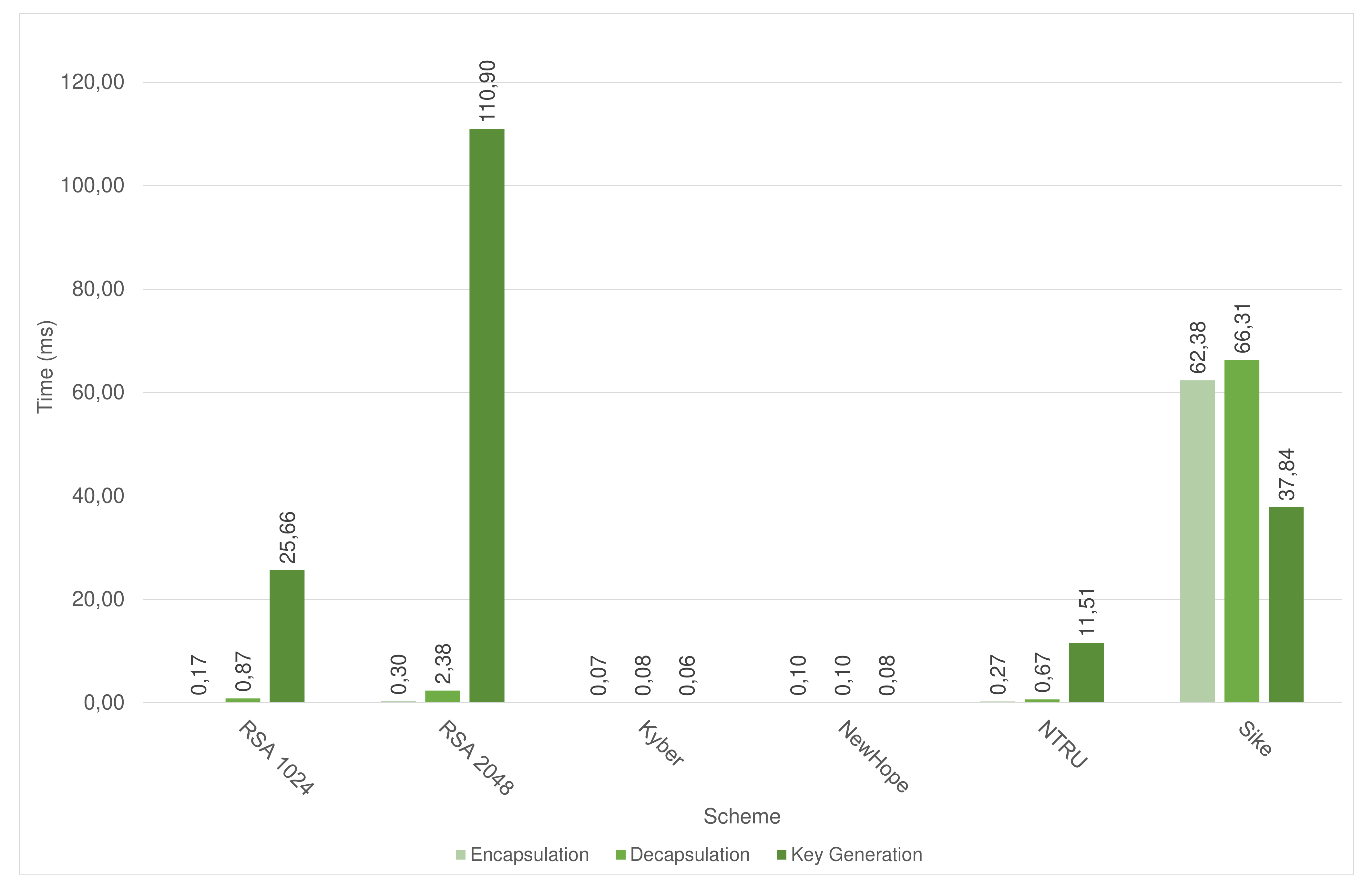}
    \caption{CPU cycles of Table~\ref{tab:cpuencap} converted to time.}
  \label{fig:cpuencapfig}
\end{figure}

We performed measurements and obtained benchmarks for the following properties of the encryption schemes:
\begin{enumerate}
	\item Public key, private key and ciphertext sizes,
	\item CPU cycles for \texttt{RSA} key generation,
	\item CPU cycles for quantum-safe key generation,
	\item CPU cycles for \texttt{RSA} encryption and decryption,
	\item CPU cycles for quantum-safe encapsulation and decapsulation.
\end{enumerate}
To get an average result for the CPU cycle measurements, \numprint{1000} iterations were run with each test, the number of CPU cycles corresponding to one second is \numprint{2399753472} cycles. Table~\ref{tab:pqkeys} contains the public key, private key and ciphertext sizes. Table~\ref{tab:cpuencap} contains the number of CPU cycles needed for encapsulation and decapsulation of a message and CPU cycles required for key generation, Figure~\ref{fig:cpuencapfig} illustrates this in time (ms) for all the schemes tested.

\begin{table}[!ht]
    \centering
    \begin{tabular}{|l|r|r|r|}
    \hline
         \textbf{Scheme} & \textbf{Public key size} & \textbf{Private key size} & \textbf{Ciphertext size} \\
         & \textbf{(bytes)} & \textbf{(bytes)} & \textbf{(bytes)} \\
         \hline
         \texttt{RSA-1024} & $<$ 128 & $<$ 128 & 128 \\
         \texttt{RSA-2048} & $<$ 256 & $<$ 256 & 256 \\
         \texttt{Frodo-640-AES}~\cite{frodo} & \numprint{9616} & \numprint{19888} & \numprint{9720} \\
         \texttt{Frodo-640-SHAKE}~\cite{frodo} & \numprint{9616} & \numprint{19888} & \numprint{9720} \\
         \texttt{Kyber512}~\cite{kyber} & 800 & \numprint{1632} & 736 \\
         \texttt{NewHope-512-CCA}~\cite{newhope} & 928 & \numprint{1888} & \numprint{1120}\\
         \texttt{NTRU-HPS-2048-509}~\cite{ntru} & 699 & 935 & 699 \\
         \texttt{Sike-p503}~\cite{sike} & 378 & 434 & 402\\
         \hline
    \end{tabular}
    \caption{The key sizes for \texttt{RSA} and quantum-safe schemes.}
    \label{tab:pqkeys}
\end{table}

Key lengths have an effect on network load as they are sent to the Directory Authorities, and the Directory Authorities distribute them to the clients. Larger ciphertexts have a significant detrimental effect on the stability, reliability and performance of a network as an increase in ciphertext size has a direct consequence on network load. We observe that, both \texttt{Frodo-640-AES} and \texttt{Frodo-640-SHAKE} have a problematic ciphertext size of \numprint{9720} bytes.

From Table \ref{tab:cpuencap}, we observe that the lattice-based quantum-safe schemes (\texttt{Kyber, NewHope, NTRU}) require less CPU cycles for generating keys than \texttt{RSA}-1024.  \texttt{Kyber} and \texttt{NewHope} drastically outperform the other schemes. We also note that the supersingular isogeny-based quantum-safe scheme \texttt{Sike}, even though slightly less performant than \texttt{RSA}-1024, outperforms \texttt{RSA}-2048. 

Key generation only affects the nodes as they generate keys based on the generation time defined in Tor. The factor that affects both the nodes and the client is the time/computation needed to encapsulate and decapsulate messages. Benchmark encapsulation and decapsulation measurements are presented in Table~\ref{tab:cpuencap}. Opposed to key generation times, where all lattice-based schemes outperformed \texttt{RSA}-1024, we observe that \texttt{NTRU} requires more CPU cycles for encapsulation. However, decapsulation for lattice-based implementations require less CPU cycles than \texttt{RSA}-1024. \texttt{Sike} requires the most CPU cycles for encapsulations and decapsulations, as it is almost 48 times more computationally heavy than \texttt{RSA}-2048.

\begin{table}[!ht]
    \begin{adjustbox}{max width=\textwidth, center}
    \begin{tabular}{|l|r|r|r|}
    \hline
         \textbf{Scheme} & \textbf{Encapsulation} & \textbf{Decapsulation} & \textbf{Key generation} \\
         \hline
         \texttt{RSA-1024} & \numprint{410402} & \numprint{2078161} & \numprint{61568194}\\
         \hline
         \texttt{RSA-2048} & \numprint{730570} & \numprint{5718858} & \numprint{266140623} \\
         \hline
         \texttt{Kyber512} & \numprint{170856} & \numprint{195106} & \numprint{152973} \\
         \hline
         \texttt{NewHope-512-CCA} & \numprint{228687} & \numprint{247457} & \numprint{193367}\\
         \hline
         \texttt{NTRU-HPS-2048-509} &  \numprint{636263} & \numprint{1609748} & \numprint{27632969}\\
         \hline
         \texttt{Sike-p503} &  \numprint{149691623} & \numprint{159119760} & \numprint{90800645} \\
         \hline
    \end{tabular}
    \end{adjustbox}
    \caption{CPU cycles for encapsulation, decapsulation and key generation averaged over \numprint{1000} test runs.}
    \label{tab:cpuencap}
\end{table}

Lattice-based schemes (\texttt{Kyber, NewHope, NTRU}) have better performance for CPU cycles than the \texttt{RSA} schemes. This suggests that these are the most fit candidates for replacing classical cryptographic schemes. However, based on ciphertext sizes, \texttt{Sike} is the best fit as the ciphertext size fits within one Tor cell (512 bytes).

\subsection{Impact}
A migration of classical cryptography to quantum-safe cryptography can have a big effect on the overall availability, reliability, stability and performance of Tor. An important factor to take into account is network load. An increase in the number of packets needed to transfer the ciphertexts of the schemes has a large effect on the network performance. 

The factor computation time, on the other end, influences the response time to users as an increase in computation time directly influences response time. The impact of computation time will be felt in the performance of Tor.

We note that a trade-off has to be made between network load and computation time, when considering the schemes that we tested.

\section{Case study: Circuit creation in Tor}
To investigate the impact of post-quantum cryptography on Tor, we propose to investigate the performance of one particular protocol, namely circuit creation. Our framework uses the SweetOnions implementation\footnote{\url{https://github.com/LeonHeTheFirst/SweetOnions}, accessed on 28/11/2019, 11:21am} which is a simplified version of the circuit creation protocol used in Tor. We consider two quantum-safe versions of this protocol: a version in which we only use post-quantum cryptography (QSO), and a hybrid version of the protocol (HSO) in which we combine the currently used cryptography with post-quantum cryptography. The reference implementation (SO) that uses standard cryptographic schemes, namely \texttt{RSA}, is also evaluated. We now provide a detailed description of each protocol. 
\subsection{Protocol descriptions}
\subsubsection{Sweet Onion (SO) protocol}

\newenvironment{protocol}[1][htb]
{\renewcommand{\algorithmcfname}{Protocol}
 \begin{algorithm}[#1]%
}{\end{algorithm}}
\begin{protocol}
\caption{Sweet Onion (SO)}
\label{prot:OS}
\centering
\trimbox{0.6cm 0cm -1.3cm 0cm}{ 
\begin{tikzpicture}
{\footnotesize
\matrix (m)[matrix of nodes, column  sep=0.5cm,row  sep=8mm, nodes={draw=none, anchor=base west,text depth=1pt, align=left} ]
{
 Client $(m)$                               &                        & Node $(pk_N)$                     \\[-6mm]
 $K_{AES} \leftarrow_R \{0,1\}^{256}$     &                        &																	 \\[-8mm]
 $c \leftarrow Enc_{AES}(K_{AES},m)$   &                        &   \\[-8mm]
 $ c' \leftarrow Enc_{RSA}(pk_N, K_{AES})$ &                        &     \\[-10mm]
                                       & $\xrightarrow{\hspace{3pt}(c,c')\hspace{5pt}}$   &  \\[-8mm]
           &    &  $K_{AES} \leftarrow Dec_{RSA}(sk_N, c') $ \\[-8mm]
           &    &  $m \leftarrow Dec_{AES}(K_{AES}, c) $  \\[-10mm]              
};
\draw[shorten <=-0.1cm,shorten >=-0.1cm] (m-1-1.south east)--(m-1-1.south west);
\draw[shorten <=-0.1cm,shorten >=-0.1cm] (m-1-3.south east)--(m-1-3.south west);
}
\end{tikzpicture}}
\end{protocol}
\vspace{-0.7cm}
In the original SweetOnions protocol, defined in Protocol \ref{prot:OS} for one layer, the client who aims to send a message $m$ to node $N$, $N\in \mathbf{N}$, encapsulates the symmetric data encryption key using $N$'s public \texttt{RSA} key $pk_N$.
To set up a circuit the client has to perform these steps with all the nodes in the circuit. Once the client knows the address of every node, this is done in sequence, each node between the client and destination decrypts one layer of encryption and forwards the message.


\vspace{-0.7cm}
\subsubsection{Quantum-safe Sweet Onion (QSO)}
\begin{protocol}
\caption{Quantum-safe Sweet Onion (QSO)}
\label{prot:QOS}
\centering
\trimbox{0.6cm 0cm -1.3cm 0cm}{ 
\begin{tikzpicture}
{\footnotesize
\matrix (m)[matrix of nodes, column  sep=0.5cm,row  sep=8mm, nodes={draw=none, anchor=base west,text depth=1pt, align=left} ]
{
 Client $(m)$                               &                        & Node $(pk^{PQ}_N)$                     \\[-6mm]
 $K_{AES} \leftarrow_R \{0,1\}^{256}$     &                        &																	 \\[-8mm]
 $c \leftarrow Enc_{AES}(K_{AES},m)$   &                        &   \\[-8mm]
 $c' \leftarrow Enc_{PQC}(pk^{PQ}_N, K_{AES})$ &                        &     \\[-10mm]
                                       & $\xrightarrow{\hspace{3pt}(c,c')\hspace{5pt}}$   &  \\[-8mm]
           &    &  $K_{AES} \leftarrow Dec_{PQC}(sk^{PQ}_N, c') $ \\[-8mm]
           &    &  $m \leftarrow Dec_{AES}(K_{AES}, c) $  \\[-10mm]             
    };
\draw[shorten <=-0.1cm,shorten >=-0.1cm] (m-1-1.south east)--(m-1-1.south west);
\draw[shorten <=-0.1cm,shorten >=-0.1cm] (m-1-3.south east)--(m-1-3.south west);
}
\end{tikzpicture}}
\end{protocol} 
\vspace{-0.7cm}
 The QSO corresponds to the simple quantum-safe variant of SO: the \texttt{RSA} key encapsulation method (KEM) is exchanged with a post-quantum KEM (PQC), see Protocol \ref{prot:QOS}. The public-private key pair of the node consists of post-quantum keys.

\subsubsection{Hybrid Sweet Onion (HSO)}


\begin{protocol}
\caption{Hybrid Sweet Onion (HSO)}
\label{prot:HOS}
\centering
\trimbox{0.6cm 0cm -1.3cm 0cm}{ 
\begin{tikzpicture}
{\footnotesize
\matrix (m)[matrix of nodes, column  sep=0.5cm,row  sep=8mm, nodes={draw=none, anchor=base west,text depth=1pt, align=left} ]
{
 Client $(m)$                               &                        & Node $(pk_N,pk^{PQ}_N)$                     \\[-6mm]
 $K^1_{AES}, K^2_{AES} \leftarrow_R \{0,1\}^{256}$     &                        &																	 \\[-8mm]
 $K_{AES} = K^1_{AES} \oplus K^2_{AES}$     &                        &																	 \\[-8mm]
 $c \leftarrow Enc_{AES}(K_{AES},m)$   &                        &   \\[-8mm]
 $c' \leftarrow Enc_{RSA}(pk_N, K^1_{AES})$ &                        &     \\[-10mm]
 $c'' \leftarrow Enc_{PQC}(pk^{PQ}_N, K^2_{AES})$ &                        &     \\[-10mm]
                                       & $\xrightarrow{\hspace{3pt}(c,c,c'')\hspace{5pt}}$   &  \\[-8mm]
           &    &  $K^1_{AES} \leftarrow Dec_{RSA}(sk_N, c') $ \\[-8mm]           
           &    &  $K^2_{AES} \leftarrow Dec_{PQC}(sk^{PQ}_N, c'') $ \\[-8mm]
           &    &   $K_{AES}=K^1_{AES} \oplus K^2_{AES} $ \\[-8mm]
           &    &  $m \leftarrow Dec_{AES}(K_{AES}, c) $  \\[-10mm]              
};
\draw[shorten <=-0.1cm,shorten >=-0.1cm] (m-1-1.south east)--(m-1-1.south west);
\draw[shorten <=-0.1cm,shorten >=-0.1cm] (m-1-3.south east)--(m-1-3.south west);
}
\end{tikzpicture}}
\end{protocol} 
 In the hybrid SweetOnions (HSO) protocol, the \texttt{RSA} KEM is combined with a post-quantum KEM. Hence, the client randomly generates two symmetric encryption keys. The first key is encapsulated with the standard \texttt{RSA} encryption algorithm and the second key is encapsulated with a post-quantum encryption algorithm. The actual data encryption key is the result of a simple XOR of these two symmetric keys. Hence the receiver should perform two decapsulation operations (one with \texttt{RSA} and one with post-quantum decryption). 

\subsection{Experimental results of quantum-safe circuit builds}
\label{circuitbuildresults}
In this section, we evaluate the performance of each protocol in terms of CPU and bandwidth consumption. The size of one Tor packet is 512 bytes. For the reference SO protocol, the underlying encryption algorithms are \texttt{RSA}-2048 and \texttt{AES}-192. For the two other protocols, the post-quantum cryptographic schemes studied in   Section \ref{sec:benchmark} are used. 

Experimental results on CPU consumption and bandwidth overhead are given in Table \ref{tab:circuitbuild}. In particular, we evaluate the cost of wrapping the layers of encryption, decapsulating one layer of encryption, and the overall circuit creation. The table also includes the size of one message and the number of packets needed for this protocol. 

\begin{table}[!htb]
  \setlength\belowcaptionskip{-5pt}
    \begin{adjustbox}{max width=\textwidth, center}
    \begin{tabular}{|l|r|r|r|r|r|r|}
    \hline
         \textbf{Scheme} & \textbf{Wrap encryp-} & \textbf{Remove} & \textbf{Total circuit} & \textbf{Message} & \textbf{Packets} & \textbf{Time}\\
         & \textbf{tion layers} & \textbf{one layer} & \textbf{ build} & \textbf{size (bytes)} & \textbf{needed} & \textbf{needed}\\
         \hline
         Original & \numprint{5131765} & \numprint{13714147} & \numprint{46274206} & \numprint{1223} & 3 & 0.0193s\\
         \hline
         \texttt{Kyber} & \numprint{1371999} & \numprint{917080} & \numprint{4123240} & \numprint{3248} & 7 & 0.0017s\\
         \hline
         \texttt{NewHope} & \numprint{1618934} & \numprint{1119668} & \numprint{4977938} & \numprint{4832} & 10 & 0.0021s\\
         \hline
         \texttt{NTRU} & \numprint{2803358} & \numprint{4149134} & \numprint{15250759} & \numprint{3099} & 7 & 0.0064s\\
         \hline
         \texttt{Sike} & \numprint{452691951} & \numprint{271667313} & \numprint{1267693889} & \numprint{1874} & \numprint{4} & 0.5283s\\
         \hline
         Hybrid \texttt{Kyber} & \numprint{6188037} & \numprint{15734659} & \numprint{53392015} & \numprint{5774} & \numprint{12} & 0.0222s\\
         \hline
        Hybrid \texttt{NewHope} & \numprint{6953196} & \numprint{13771785} & \numprint{48268550} & \numprint{7886} & \numprint{16} & 0.0201s\\
         \hline
        Hybrid \texttt{NTRU} & \numprint{7517316} & \numprint{18977229} & \numprint{64449002} & \numprint{5550} & \numprint{11} & 0.0269s\\
         \hline
        Hybrid \texttt{Sike} & \numprint{456441243} & \numprint{275867016} & \numprint{1284042291} & \numprint{3938} & \numprint{8} & 0.5351s\\
         \hline
    \end{tabular}
    \end{adjustbox}
    \caption{The CPU cycles needed for building a circuit (averaged over \numprint{1000} test runs) and message sizes.}
    \label{tab:circuitbuild}
\end{table}

\subsubsection{Quantum-safe Sweet Onion (QSO) experimental results}
\label{quantumsaferesults}




We observe that QSO based on all lattice-based post-quantum schemes outperforms the original SO. On the other hand, while the integration of \texttt{Sike} increases the overall time significantly, the bandwidth overhead is very close to SO.


To summarize, a lattice-based scheme may be considered as a potential cryptographic primitive for circuit building since these schemes require less CPU cycles compared to \texttt{Sike}. Nevertheless, the use of lattice-based schemes significantly increases the number of packets and network load compared to \texttt{Sike}. Therefore, depending on the original communication cost, one can decide whether to choose \texttt{Sike} or a lattice-based PQC. 

\subsubsection{Hybrid Sweet Onion (HSO) experimental results}
\label{combinedresults}
When using the hybrid scheme, we observe that both the computational cost and the bandwidth  increases significantly. This is mainly due to the fact that HSO uses one \texttt{RSA} encapsulation and one encapsulation with PQC. Consequently, the cost originating from PQC for lattice-based cryptographic schemes becomes negligible when combined with \texttt{RSA}. 
Even though CPU consumption remains affordable in the hybrid implementation, the bandwidth overhead is important. The number of packets is at least doubled when switching to the hybrid solutions. 

\section{Conclusion}
\label{Conc}

In this paper, we investigated the main challenges to develop a quantum-safe Tor network and focused on the algorithms that use long-term and medium-term keys. Experimental studies show that among the six post-quantum cryptographic scheme evaluated, there is no single winning solution. Nevertheless, given the current status of the NIST standardisation process, \texttt{Sike} seems the most optimal one when it comes to assessing the communication overhead. 

As for future work, it may be interesting to test other schemes such as the code-based \texttt{BIKE}. Testing the remaining lattice and isogeny-based schemes is also an interesting future topic as they might have better performance measurements than the ones currently available in the Open Quantum Safe library.
As for field experiments, an implementation of Tor called \texttt{TorLAB}~\cite{torlab} is available and simulates Tor on a private network of Raspberry PIs. It would be beneficial to re-create the network and extend the measurements of our research to the network load. This would ensure a more realistic study for the evaluation of expected circuit build times, since in the current setting, network latency is omitted.

\bibliographystyle{splncs04}
\bibliography{references}
\end{document}